# A STUDY OF EVOLUTION OF NEAR-EARTH DAEMON'S FLUXES WITH USING DARK ELECTRON MULTIPLIERS (DEMs)

E.M. DROBYSHEVSKI*, M.E. DROBYSHEVSKI**, S.A. PONYAEV***

*Ioffe Physical-Technical Institute, Russian Academy of Sciences, 194021 St-Petersburg, Russia*
*\*emdrob@mail.ioffe.ru, \*\*miked@mail.ru, \*\*\*serguei.poniaev@mail.ioffe.ru*

DEMs have been used to experimental studying the temporal evolution of the March maximum of fluxes of near-Earth daemons. It is shown that part of objects from near-Earth almost circular heliocentric orbits (NEACHOs), from which a rather intense flux proceeds during only about four weeks, forms in the second half of March the population in geocentric Earth-surface-crossing orbits (GESCOs). The resistance of the Earth's matter results in that GESCO objects sink into the Earth's interior, so that the GESCO population nearly disappears by the end of April.

*Keywords*: DM detection; Planck black holes; DM objects' interaction with matter

PACS Nos.: 95.35.+d

## 1. Daemons and DEMs

Judging from the presently available data, daemons (Dark Electric Matter Objects) are Planck elementary black holes with $m_P \approx 3 \times 10^{-5}$ g at $r_g \approx 2 \times 10^{-33}$ cm. They can carry an electric charge of up to $Ze \approx 10e$. Negatively charged objects can be detected because of capturing atomic nuclei in a substance (charge $Z_n$, atomic mass $A_n$), with release of a binding energy $W \approx 1.8 Z Z_n A_n^{-1/3}$ MeV (~100 MeV for Fe or Zn) [1]. An excited nucleus emits scintillation-active particles (nucleons and complexes of these, electrons, etc.) and the nucleus remainder is carried away by the supermassive daemon. Residing in the nucleons of the captured nucleus remnant, daemon catalyzes their successive decay, so that, in the end, the complex formed by a daemon and the nucleus remainder and, possibly, by other atoms chemically bound to this nucleus (c-daemon) acquires, first, zero and then, even a negative charge, which enables capture of a new nucleus, with subsequent repetition of the process.

We are interested in objects somehow falling onto the Earth or passing through the planet. It can be shown that some daemons from the galactic disk (with a characteristic velocity dispersion ~4÷20 km/s [2]) are decelerated when passing through the Sun to such an extent that they are captured by the star to strongly elongated orbits that gradually shrink into the Sun. If an object on such an orbit passes through the Earth's sphere of action, its perihelion moves out of the Sun's interior with high probability and the object transfers into a strongly elongated Earth-crossing heliocentric orbit (SEECHO), which is stable until the next passage of the object near the Earth. Gravitational interactions of SEECHO daemons with the Earth transfer some of these to near-Earth almost circular heliocentric orbits (NEACHOs). It follows from the celestial mechanics of these processes that [3] objects accumulated on SEECHOs pass close to the Earth's orbit mostly somewhere near the beginning of June, whereas NEACHOs in their perihelia touch the Earth's orbit mostly a week before the equinoxes. Daemons strike the Earth with $V \approx$ 10-15 km/s from NEACHOs and $V \approx$ 30-50 km/s from SEECHOs. The latter qualitatively accounts for the results of DAMA experiments at Gran Sasso and CoGeNT at Soudan [4]. The passage of NEACHO daemons through the Earth's body diminishes the velocity of some of these to such an extent that they transfer into geocentric Earth-surface-crossing orbits (GESCOs) with a perigees within the Earth. The resistance of the Earth's matter leads to a gradual shrinking of the GESCOs. As a result, a daemon kernel is formed at the Earth's center, which accounts for a multitude of geophysical facts [5].

Our experiments with a thin (~10 μm) layer of a ZnS(Ag) scintillator did reveal the incidence of objects with $V \sim$ 10-15 km/s in March and September [6]. These objects excite in ZnS(Ag) prolonged scintillations characteristic of heavy non-relativistic particles of the type of α-particles (HPS, heavy particle scintillations) [1]. In addition, it was found that daemons passing through some photomultiplier tubes (we used a 5" PMT of the FEU-167 type) excite directly in the tubes short signals resembling in shape intrinsic PMT noises [7] (or signals excited by cosmic-ray air showers, which are actually signals from scintillations in the glass of the PMT bulb [8]). These are noise-like signals (NLSs).

The excitation of NLSs by a c-daemon passing through a PMT is quite understandable because the c-daemon is an actively living object with permanently transforming constituents, both the remainder of a captured nucleus with its electron shells, carried by the daemon, and, possibly, a chemically bound complex of other atoms (hence follows that the environment of a detector: blocks of cement, lead, copper, etc., can affect measurement results). Indeed, absolutely light-insulated dark electron multipliers (DEMs) specially fabricated from FEU-167, with the inner surface of the near-cathode section of the glass bulb fully covered with an Al layer proved to be good in daemon detection. Of particular interest is a variant with increased (to ~0.5 μm) thickness of the Al coating on the inner surface of a planar front disk. TEU-167d multipliers of this kind detect the direction in which c-daemons pass through the devices, mostly from the vacuum within the bulb into the thickened Al coating on the front disk [9]. Presently, the main obstacle to daemon detection by DEMs is presented by the outside electromagnetic noise whose signals are frequently indistinguishable in shape from the valid signal.

For this purpose, we performed, with assistance of well-disposed staff members of the Central Astronomical Observatory, Russian Academy of Sciences, an experiment at Pulkovo outside the St. Petersburg city limits. Although the electromagnetic noise from the nearby Pulkovo airport was rather strong, this one-week (March 19-26, 2011) experiment nevertheless yielded positive results and revealed a number of interesting points [4].

First, raising to a certain extent the thickness of the ZnS(Ag) layer deposited onto the upper surface of a horizontal polystyrene plate and making higher the feed voltage of the DEM improved the overall sensitivity of the detector: a flux $f \approx 8 \times 10^{-7}$ cm$^{-2}$ s$^{-1}$ was measured, which somewhat exceeds the values measured in the past.

Second, it was found that the DEM with a downwards-oriented screen (channel 21, TEU-167d no. 00159) records a nearly doubled flux of daemons (from above) than TEU-167d (no. 00160) with an upward-oriented screen (channel 2), which is assumed to favor recording by this DEM of objects passing through the Earth from below upwards.

We believe that there are two most probable reasons for this behavior: differences in the device (and even the module in total) design and(or) those in the trajectory directions. For most of primary objects incident from above, the trajectories are nearly vertical, whereas the trajectories of daemons passing through the Earth may be noticeably inclined to the vertical direction, because of which an object does not pass through both the detecting elements (ZnS(Ag) layer and DEM) at angles $\geq \pi/4$ and, therefore, is not recorded.

In the present study, we interchanged both these DEMs, made their voltages normal, and reduced the thickness of the ZnS(Ag) layer on the upper surface of the polystyrene plate to the normal value (~3.5 mg/cm$^2$).

The exposure of the detector at the Ioffe Institute in St. Petersburg was lengthened from one to nine weeks (from February 26 to April 28, 2012). As a result, we revealed some specific features of the temporal evolution of fluxes of daemons that accumulated on NEACHOs and then passed to GESCOs.

**2. Experimental**

Now, TEU-167d no. 00159 had its front screen directed upwards and was placed at the bottom of the detector module. Together with FEU-167 observing from above the ZnS(Ag) screen whose HPSs induced a pulse in FEU-167 and triggered the first traces of two S9-8 oscilloscopes, this DEM no. 00159 constituted channel 3. The screen of TEU-167d no. 00160 placed in a cubic (with side of 51 cm) tinned-iron case was now oriented downwards. TEU-167d no. 00160 formed, together with FEU-167, channel 23. The horizontally oriented and facing each other screens of TEU-167d nos. 00159 and 00160 were separated by a thin (~4 mm) light-tight dielectric spacer (it will be recalled that the TEU-167d face screens had in their central part a transparent window, ~11 mm in diameter, in the Al coating for rough light calibration of the devices). These screens were 29 cm below the ZnS(Ag) plane. Signals from these two TEU-167d were delivered to the second traces of the corresponding oscilloscopes. If signals were present in both traces of any of the oscilloscopes, the event was memorized by a personal computer. The design of the detector unit containing a polystyrene plate with a ZnS(Ag) layer on its upper surface, observed from above by FEU-167, and two TEU-167d with nearly contacting screens was shown in Fig. 2 in [9].

The results of our experiment for channels 23 and 3 are presented in Fig. 1 and in Table 1 also. We processed by the standard procedure only the events with a HPS in the first (triggering) trace of the oscilloscope. (It should be noted at once that the number of events in the $+90 \leq \Delta t \leq 100$ μs bin may be somewhat overstated because a considerable part of signal "tails" fall outside the oscilloscope screen. Therefore, their real shape is difficult to judge and they quite may be noise-related. It will also be recalled that events with $\Delta t \leq 0.4$ μs were normally dropped because of possibly being caused by relativistic cosmic-ray particles.)

The number of events in each week was chosen by selecting the signal amplitude from TEU-167d (at $U_2 \geq 0.4 \div 0.8$ mV) so that their total number was within the range 100-250. The reason for this selection is in the discrete (0.2 mV) digitization of the signal amplitude by the oscilloscope (see, e.g. data on 4÷10 March and 1÷7 April for the channel 3).

**3. Basic Results**

Primarily noteworthy is the fact that channel 3 having an upward-directed screen (this is TEU-167d no. 00159), which should have seemingly favored predominant recording of daemons coming from below (i.e., events with $\Delta t < 0$), reveals no anyhow attractive features corresponding to $V \approx -10 \div -15$ km/s in its $N(\Delta t)$ distribution. A similar circumstance has been noted in the Pulkovo experiment [4]. This may indicate that the expected flux from below is indeed not recorded properly because of the strong inclination of trajectories of objects coming from below, say. However, it worth noting the lack of the top iron cover on the tinned iron cube of the module imbedding the scintillator and the channel 23 with the TEU-167d no. 00160; the top cover is made of a black paper [1]. As the same concerns the Pulkovo experiment also [4] one has to conclude that the phenomenon of a greater sensibility of the intra-cube TEU-167d is not caused by the from-bottom-to-top trajectories' peculiarities but it is caused by neighboring material composition, seemingly. Indeed, some daemons going from below through the teen iron lower cover can capture the massive atomic nuclei here and so become for a time incapable of capturing a new nucleus (say, in the ZnS(Ag) scintillator).

Table 1. Per week $N(\Delta t)$ distributions for the pair events triggered by the HPS in the first trace of an oscilloscope for channels 23 and 3. $U_2$ is the lower level of NLS in the second oscilloscopic trace. The bins, each 20 μs wide, are centered at the $\Delta t$ specified in the Table. $\sum$ is the evens' number in left and right wings of the $N(\Delta t)$. One can see the continuous shift of the maxima in the channel 23 $N(\Delta t)$ right wing from +30 μs to +70 μs bin during these nine weeks (see the bold numbers; for the first four weeks the +30 μs maximum has the C.L. = $3\sigma$, see the text). Such a shift corresponds to a gradual drop of velocity from 10-15 km/s (NEACHO objects) to about 4-5 km/s (GESCO objects). No noteworthy trend was found in other parts of $N(\Delta t)$'s. The data for 04.03-10.03 and 01.04-07.04, 2012 (channel 3) illustrate an influence of the oscilloscopic digitization on the $N(\Delta t)$ parameters.

| Channel 23 | | | | | | | | | | | | |
|---|---|---|---|---|---|---|---|---|---|---|---|---|
| Dates, 2012 | $U_2$, mV | -90 | -70 | -50 | -30 | -10 | $\sum_-$ | +10 | +30 | +50 | +70 | +90 | $\sum_+$ |
| 26.02-03.03 | 0.6 | 12 | 15 | 17 | 18 | 13 | 75 | 15 | **18** | 14 | 21 | 7 | 75 |
| 04.03-10.03 | 0.6 | 15 | 22 | 19 | 12 | 15 | 83 | 18 | **17** | 15 | 10 | 16 | 76 |
| 11.03-17.03 | 0.6 | 10 | 9 | 12 | 13 | 4 | 48 | 7 | **17** | 9 | 11 | 10 | 54 |
| 18.03-24.03 | 0.6 | 14 | 9 | 14 | 10 | 13 | 60 | 12 | **16** | 16 | 12 | 9 | 65 |
| 25.03-31.03 | 0.6 | 10 | 18 | 14 | 8 | 16 | 66 | 20 | 8 | **12** | 10 | 10 | 60 |
| 01.04-07.04 | 0.6 | 18 | 17 | 21 | 18 | 14 | 88 | 17 | 14 | **21** | 14 | 22 | 88 |
| 08.04-14.04 | 0.6 | 11 | 11 | 13 | 15 | 12 | 62 | 15 | 14 | **24** | 21 | 22 | 96 |
| 15.04-21.04 | 0.6 | 17 | 27 | 23 | 19 | 19 | 105 | 22 | 16 | **20** | **20** | 21 | 99 |
| 22.04-28.04 | 0.6 | 18 | 26 | 28 | 23 | 25 | 120 | 27 | 26 | **31** | **31** | 24 | 139 |
| Channel 3 | | | | | | | | | | | | |
| Dates, 2012 | $U_2$, mV | -90 | -70 | -50 | -30 | -10 | $\sum_-$ | +10 | +30 | +50 | +70 | +90 | $\sum_+$ |
| 26.02-03.03 | 1.0 | 15 | 9 | 13 | 12 | 16 | 65 | 16 | 23 | 6 | 9 | 21 | 75 |
| 04.03-10.03 | 0.8 | 21 | 26 | 26 | 16 | 21 | 110 | 13 | 18 | 24 | 19 | 26 | 100 |
| 04.03-10.03 | 1.0 | 7 | 12 | 8 | 6 | 0 | 33 | 1 | 7 | 11 | 8 | 8 | 35 |
| 11.03-17.03 | 0.6 | 15 | 30 | 13 | 13 | 28 | 99 | 14 | 23 | 18 | 18 | 31 | 104 |
| 18.03-24.03 | 0.6 | 21 | 19 | 18 | 24 | 19 | 101 | 16 | 16 | 23 | 19 | 18 | 92 |
| 25.03-31.03 | 0.6 | 29 | 25 | 23 | 22 | 18 | 117 | 27 | 35 | 32 | 23 | 31 | 148 |
| 01.04-07.04 | 0.6 | 11 | 12 | 6 | 12 | 16 | 57 | 14 | 13 | 6 | 14 | 17 | 64 |
| 01.04-07.04 | 0.4 | 16 | 20 | 12 | 16 | 20 | 84 | 22 | 17 | 14 | 17 | 27 | 97 |
| 08.04-14.04 | 0.6 | 14 | 15 | 19 | 22 | 18 | 88 | 12 | 12 | 11 | 15 | 22 | 72 |
| 15.04-21.04 | 0.6 | 12 | 13 | 13 | 12 | 12 | 62 | 17 | 15 | 13 | 11 | 16 | 72 |
| 22.04-28.04 | 0.6 | 13 | 19 | 18 | 17 | 14 | 81 | 16 | 13 | 15 | 15 | 22 | 81 |

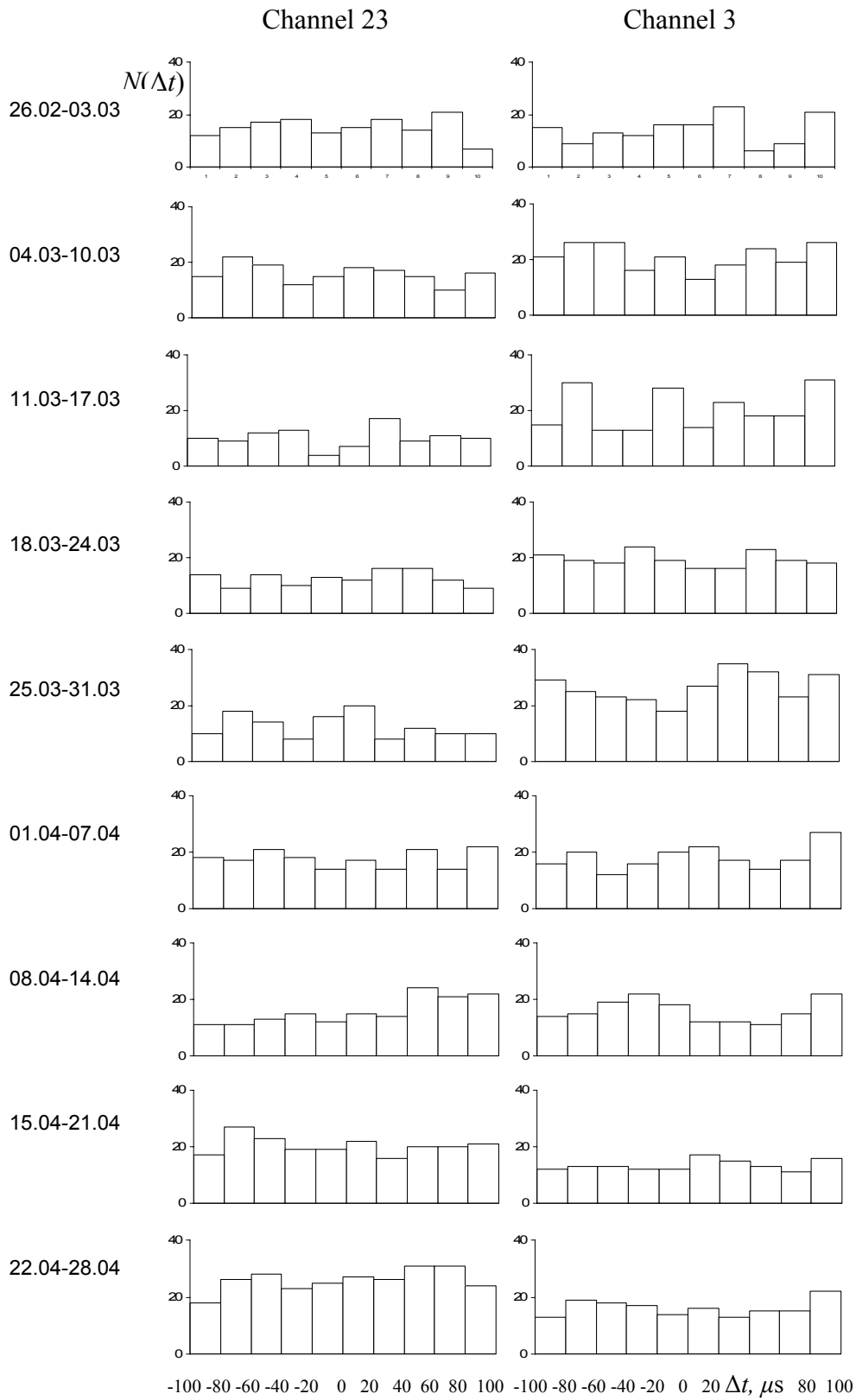

Fig.1. Graphic presentation of $N(\Delta t)$ dependence on time during the nine weeks in March and April, 2012, for channels 23 and 3.

As expected in terms of the concepts we develop, channel 23 with DEM TEU-167d no. 00160 having a downward-directed front screen records the flux incident from above (see Table 1 and Fig. 1). This flux is primarily manifested in the $20 < \Delta t < 40$ μs bin (NEACHO objects). Its reliability (confidence level) during four weeks (February 26 through March 24) is $3\sigma$, a rather high value for measurements with a single unit (if one combines these results with the earlier data [10], the C.L. becomes $5.4\sigma$). However, after March 18, the flux of NEACHO objects becomes weaker and indications of existence of a flux of daemons shift to the $40 < \Delta t < 60$ μs bin (5 km/s $< V <$ 7.5 km/s) corresponding to GESCO objects appear (see also [9]). Beginning in mid-April, this flux becomes weaker too, because, possibly, most part of objects captured from NEACHOs to GESCOs entered the Earth's interior.

**4. Conclusions**

DEMs rather effectively detect daemons. However, as any newly appeared device, they need an improvement, which requires more in-depth studies of the interaction of daemons with matter (for a start, at least consideration of theoretical aspects of how supermassive particles with variable electric charge behave and move in substance).

A larger number of events is recorded by channels with DEMs enclosed in the module teen iron box and whose screens look downwards. Possibly, it is caused by the detector asymmetry as the top cover of the module box is the black paper. But if the upper DEM records more events and this could presumably due also to differences between the trajectory inclinations for objects coming from above and below, it is not improbable that the specific feature under discussion may depend on the latitude at which a detector is situated.

The hypothesis that a part of the NEACHO flux passes to GESCO objects was confirmed. Indications of this transition are most clearly manifested in the third ten-day period of March.

The flux of daemons, especially that of GESCO objects, decreases beginning at mid-April. Thus, the main evolution of spring fluxes of NEACHO-GESCO objects occurs during eight to nine weeks, in agreement with the previous estimate of two to three months [1,6].

**Acknowledgments**

The authors are greatly indebted to R.O. Kurakin and P.B. Simonov for a help in conducting the experiment.